\documentclass[aps,prd,onecolumn,groupedaddress,showpacs,nofootinbib,amssymb]{revtex4}
\usepackage[dvips]{graphicx}
\usepackage{amssymb}
\usepackage{amsmath}
\usepackage{graphicx,,color}
\usepackage{amsfonts}
\usepackage{bm}
\usepackage{cancel}
\usepackage{comment}
\newcommand\be{\begin{equation}}
\newcommand\ee{\end{equation}}

\allowdisplaybreaks[4]

\begin{document}

\title{The Early-time Cosmology with Stiff Era from Modified Gravity}
\author{S.~D. Odintsov,$^{1,2}$\,\thanks{odintsov@ieec.uab.es}
V.K. Oikonomou,$^{3,4}$\,\thanks{v.k.oikonomou1979@gmail.com}}

\affiliation{$^{1)}$ ICREA, Passeig Luis Companys, 23, 08010 Barcelona, Spain\\
$^{2)}$ Institute of Space Sciences (IEEC-CSIC) C. Can Magrans s/n,
08193 Barcelona, Spain\\
$^{3)}$ Laboratory for Theoretical Cosmology, Tomsk State University
of Control Systems
and Radioelectronics, 634050 Tomsk, Russia (TUSUR)\\
$^{4)}$ Tomsk State Pedagogical University, 634061 Tomsk, Russia\\
}

\tolerance=5000

\begin{abstract}
In this work, we shall incorporate a stiff era in the Universe's
evolution in the context of $F(R)$ gravity. After deriving the
vacuum $F(R)$ gravity, which may realize a stiff evolution, we
combine the stiff $F(R)$ gravity with an $R^2$ model, and we
construct a qualitative model for the inflationary and stiff era,
with the latter commencing after the end of the inflationary era. We
assume that the baryogenesis occurs during the stiff era, and we
calculate the baryon to entropy ratio, which effectively constraints
the functional form of the stiff $F(R)$ gravity. Further constraints
on the stiff $F(R)$ gravity may come from the primordial
gravitational waves, and particularly their scalar mode, which is
characteristic of the $F(R)$ gravity theory. The stiff era presence
does not contradict the standard cosmology era, namely, inflation,
and the radiation-matter domination eras. Furthermore, we
investigate which $F(R)$ gravity may realize a dust and stiff matter
dominated Einstein-Hilbert evolution.
\end{abstract}

%PACS numbers: 04.50.Kd, 95.36.+x, 98.80.-k, 98.80.Cq

\maketitle

\section{Introduction \label{SecI}}

Unquestionably, the era that follows the end of inflation up to the
nucleosynthesis epoch, is a mysterious era, for which no concrete
description exists up to date. It is believed that immediately after
the end of the inflationary era, the Universe enters the reheating
era, which heats up the cold and large Universe that results after
the exponentially fast expansion during the inflationary epoch.
Hence, it is highly likely that the epoch after inflation and before
the nucleosynthesis is radiation dominated.

However, there exist alternative hypotheses in the literature for
the post-inflationary and pre-nucleosynthesis epoch, which assume
that the Universe does not enter the radiation domination epoch
directly, but a stiff matter era precedes the radiation epoch
\cite{Zeldovich:1972zz,Barrow:1982ei,Kamionkowski:1990ni,Spokoiny:1993kt,Joyce:1997fc,Joyce:1996cp,Salati:2002md,OliveiraNeto:2011sm,Chavanis:2014lra,Chavanis:2014hba,Dariescu:2016noi,KalyanaRama:2009px}.
This stiff matter era hypothesis was introduced by Zel'dovich some
time ago \cite{Zeldovich:1972zz}, who assumed that the Universe
matter content consisted of a cold baryon gas with stiff equation of
state. Many proposals exist in the literature for the stiff matter
era, in the context of which, the energy density scales as $\rho\sim
a^{-6}$, where $a$ is the scale factor\footnote{A scaling $\rho\sim
a^{-6}$, has also been considered in non-singular dilaton cosmology
\cite{Brandenberger:1998zs} and also it has been related to
inhomogeneities \cite{Cai:2013vm}.}. In effect, the Universe expands
in a more rapid way, in comparison to a radiation dominated way, in
which case $\rho \sim a^{-4}$. In many contexts, the stiff era epoch
can be generated by a canonical scalar field, which slow-rolls down
a potential, and at the end of the slow-roll era, the kination era
commences \cite{Spokoiny:1993kt,Joyce:1997fc,Joyce:1996cp}, which is
characterized by a stiff scaling $a^{-6}$. The so-called kination
\cite{Spokoiny:1993kt,Joyce:1997fc,Joyce:1996cp} era may generate a
rapid expansion of the Universe, prior to nucleosynthesis, without
disrupting the physical processes of the latter. One of the
interesting features of the kination era is that the electroweak
baryogenesis picture is altered, in comparison to radiation
dominated models. Also, it is possible that the reheating process is
generated by high frequency gravitons, which eventually thermalize
the vast and cold post-inflationary Universe
\cite{Giovannini:1998bp}. In close connection to the graviton
reheating, a stiff era may directly affect the tensorial modes of
the primordial gravitational waves \cite{Giovannini:1998bp}.

In this work we aim to incorporate a stiff era in the Universe's
evolution, in the context of $F(R)$ gravity cosmology
\cite{reviews1,reviews2,reviews3,reviews4,reviews5}. Particularly,
by using well-known reconstruction techniques
\cite{Nojiri:2006gh,Nojiri:2009kx}, we shall investigate which
$F(R)$ gravity can generate the stiff cosmological era. By using the
resulting $F(R)$ gravity, we shall study an $F(R)$ gravity model,
which at early times is described by the $R^2$ inflationary model,
and after the inflationary era, it is described by the stiff $F(R)$
gravity. The observational constraints on the free parameters of the
$R^2$ inflationary era, and the free parameters of the stiff $F(R)$
gravity, may be chosen in such a way so that the stiff era commences
after the end of the inflationary era. This picture however can be
changed in general, since the free parameters can be chosen in an
alternative way. We shall study the first scenario, in which case
the stiff era commences at $\sim 10^{-13}$sec, however an
interesting alternative scenario is to choose the stiff era to occur
at $t\sim 1$sec, that is, prior to nucleosynthesis. In the modified
$R^2$ model, the stiff $F(R)$ gravity dominates the expansion at
approximately $\sim 10^{-13}$sec, and after that, the Universe
evolves in a more rapid way, in comparison to a radiation dominated
way. Also we shall address the baryogenesis issue, which we will
assume that it occurs during the stiff era. The mechanism that will
generate an non-zero baryon to entropy ratio is the gravitational
baryogenesis mechanism
\cite{Davoudiasl:2004gf,Lambiase:2006dq,Odintsov:2016hgc,Oikonomou:2015qfh,
Saaidi:2010ey,Sadjadi:2007dx,Lambiase:2006ft,Bento:2005xk,
Feng:2004mq,Oikonomou:2016jjh,Oikonomou:2016pnq,
Odintsov:2016apy,Arbuzova:2016cem,Lima:2016cbh,Arbuzova:2017vdj,Oikonomou:2017qyp},
and we shall study the implications of this mechanism on the stiff
era. As we will show, even in the Einstein-Hilbert case, if the
baryon asymmetry is generated during the stiff era, a non-zero
baryon to entropy ratio is predicted, in contrast to the radiation
case, which results to a zero  baryon to entropy ratio. By using the
observational bounds, we will study the restrictions imposed on the
free parameters of the stiff $F(R)$ gravity. Accordingly, we will
also investigate how the observational data on primordial
gravitational waves can restrict the stiff $F(R)$ gravity. Finally,
we shall investigate which vacuum $F(R)$ gravity can generate an
Einstein-Hilbert stiff and dust matter cosmological evolution. As an
overall comment, with this paper we demonstrate that the existence
of a stiff era following the inflationary era, may comply with
current observational bounds.

This paper is organized as follows: In section II, we shall
investigate which vacuum $F(R)$ gravity can generate a stiff matter
era. In section III we shall present a phenomenological model, which
combines an early $R^2$ inflation, followed by a stiff era. We
discuss the details of the model, and we investigate how the choice
of the free parameters affects the phenomenology of the model. Also,
we shall study the implications of the gravitational baryogenesis
mechanism on the stiff $F(R)$ gravity, and in section IV we examine
how the observational constraints on the primordial gravitational
waves, may affect the functional form of the stiff $F(R)$ gravity.
Finally, in section V, we investigate which vacuum $F(R)$ gravity
may generate an Einstein-Hilbert dust and stiff matter evolution.

\section{Vacuum $F(R)$ Gravity Description of a Stiff Era Evolution}

Our first goal is to investigate which vacuum $F(R)$ gravity can
produce the stiff matter era evolution. In the context of a
scalar-tensor gravity, a stiff matter era is generated by the
presence of a perfect fluid with equation of state $p=\rho$ and the
energy density as a function of the scale factor behaves as
$\rho\sim 1/a^{6}$. Then by solving the resulting cosmological
equations, the scale factor and the corresponding Hubble rate, that
describe the stiff era evolution are,
\begin{equation}\label{stiffevo}
a(t)=\zeta t^{\frac{1}{3}},\,\,\,H(t)=\frac{1}{3t}\, .
\end{equation}
By using a very well known reconstruction technique, in this section
we shall find the $F(R)$ gravity which generates the stiff matter
era evolution of Eq. (\ref{stiffevo}).

Before we get into the core of the calculation, let us present in
brief some essential features of $F(R)$ gravity, in order to
maintain the article self-contained. For details, see the reviews
\cite{reviews1,reviews2,reviews3,reviews4,reviews5}. We shall assume
that the geometric background is that of a
Friedmann-Robertson-Walker, with line element,
\begin{equation}
\label{metricfrw} ds^2 = - dt^2 + a(t)^2 \sum_{i=1,2,3}
\left(dx^i\right)^2\, .
\end{equation}
Also, the connection is assumed to be a metric compatible affine
connection, the Levi-Civita connection. The corresponding
Christoffel symbols are,
\begin{equation}\label{christofell}
\Gamma_{\mu \nu }^k=\frac{1}{2}g^{k\lambda }(\partial_{\mu
}g_{\lambda \nu}+\partial_{\nu }g_{\lambda \mu}-\partial_{\lambda
}g_{\mu \nu})
\end{equation}
and in addition, the Ricci scalar is equal to,
\begin{equation}\label{ricciscalar}
R=g^{\mu \nu }(\partial_{\lambda }\Gamma_{\mu \nu
}^{\lambda}-\partial_{\nu }\Gamma_{\mu \rho }^{\rho}-\Gamma_{\sigma
\nu }^{\sigma}\Gamma_{\mu \lambda }^{\sigma}+\Gamma_{\mu \rho
}^{\rho}g^{\mu \nu}\Gamma_{\mu \nu }^{\sigma}).
\end{equation}
The vacuum Jordan frame $F(R)$ gravity action is,
\begin{equation}\label{action}
\mathcal{S}=\frac{1}{2\kappa^2}\int \mathrm{d}^4x\sqrt{-g}F(R)\, ,
\end{equation}
where $\kappa^2=8\pi G=\frac{1}{M_p^2}$ and $M_p$ is the Planck mass
scale. We shall work in the context of the metric formalism, so the
equations of motion are obtained by varying the action with respect
to the metric $g_{\mu \nu}$, so the resulting equations of motion
are,
\begin{equation}\label{eqnmotion}
F'(R)R_{\mu \nu}(g)-\frac{1}{2}F(R)g_{\mu
\nu}-\nabla_{\mu}\nabla_{\nu}F'(R)+g_{\mu \nu}\square F'(R)=0\, ,
\end{equation}
or equivalently,
\begin{align}\label{modifiedeinsteineqns}
R_{\mu \nu}-\frac{1}{2}Rg_{\mu
\nu}=\frac{\kappa^2}{F'(R)}\Big{(}T_{\mu
\nu}+\frac{1}{\kappa^2}\Big{(}\frac{F(R)-RF'(R)}{2}g_{\mu
\nu}+\nabla_{\mu}\nabla_{\nu}F'(R)-g_{\mu \nu}\square
F'(R)\Big{)}\Big{)}\, ,
\end{align}
where the prime denotes differentiation with respect to the Ricci
scalar. For the metric (\ref{metricfrw}), the cosmological equations
become,
\begin{align}
\label{JGRG15} 0 =& -\frac{F(R)}{2} + 3\left(H^2 + \dot H\right)
F'(R) - 18 \left( 4H^2 \dot H + H \ddot H\right) F''(R)\, ,\\
\label{Cr4b} 0 =& \frac{F(R)}{2} - \left(\dot H + 3H^2\right)F'(R) +
6 \left( 8H^2 \dot H + 4 {\dot H}^2 + 6 H \ddot H + \dddot H\right)
F''(R) + 36\left( 4H\dot H + \ddot H\right)^2 F'''(R) \, ,
\end{align}
where $H$ stands for the Hubble rate $H=\dot a/a$ and the Ricci
scalar is $R=12H^2 + 6\dot H$. The $F(R)$ theory modifies the left
hand side of the Einstein equations, and the effects of $F(R)$
gravity can mimic the effects of a perfect fluid, as we now show.
Indeed, by introducing the effective energy density
$\rho_\mathrm{eff}$ and the effective pressure $p_\mathrm{matter}$,
\begin{align}
\label{Cr4} \rho_\mathrm{eff} =&
\frac{1}{\kappa^2}\left(-\frac{1}{2}F(R) + 3\left(H^2 + \dot
H\right) F'(R) - 18 \left(4H^2 \dot H
+ H \ddot H\right)F''(R)\right) \, ,\\
\label{Cr4bb} p_\mathrm{eff} =&
\frac{1}{\kappa^2}\left(\frac{1}{2}F(R) - \left(3H^2 + \dot H
\right)F'(R) + 6 \left(8H^2 \dot H + 4{\dot H}^2 + 6 H \ddot H +
\dddot H \right)F''(R) + 36\left(4H\dot H + \ddot H\right)^2F'''(R)
\right)\, ,
\end{align}
the cosmological equations (\ref{JGRG15}) can be written in the
Einstein-like form,
\begin{equation}
\label{JGRG11B} \rho_\mathrm{eff}=\frac{3}{\kappa^2}H^2 \, , \quad
p_\mathrm{eff}= - \frac{1}{\kappa^2}\left(3H^2 + 2\dot H\right)\, .
\end{equation}
The first of the above equations will prove quite useful in the
gravitational baryogenesis section later on.

Now, using the above equations, we shall use the reconstruction
method of Ref. \cite{Nojiri:2009kx}, and we shall investigate which
vacuum $F(R)$ gravity can generate the evolution (\ref{stiffevo}).
To this end, we rewrite the first FRW equation in
(\ref{modifiedeinsteineqns}) as follows,
\begin{equation}\label{frwf1}
-18\left ( 4H(t)^2\dot{H}(t)+H(t)\ddot{H}(t)\right )F''(R)+3\left
(H^2(t)+\dot{H}(t) \right )F'(R)-\frac{F(R)}{2}=0
\end{equation}
where $F'(R)=\frac{\mathrm{d}F(R)}{\mathrm{d}R}$. We shall introduce
the $e$-foldings number $N$ defined as follows,
\begin{equation}\label{efoldpoar}
e^{-N}=\frac{a_0}{a}\, ,
\end{equation}
and the first FRW equation in Eq. (\ref{frwf1}), can be cast in
terms of the $e$-foldings number, in the following way,
\begin{align}\label{newfrw1}
& -18\left ( 4H^3(N)H'(N)+H^2(N)(H')^2+H^3(N)H''(N) \right )F''(R)
\\ \notag & +3\left (H^2(N)+H(N)H'(N) \right
)F'(R)-\frac{F(R)}{2}=0\, .
\end{align}
Notice now that the Hubble rate is a function of the $e$-foldings
number, and the prime this time denotes differentiation with respect
to $N$, that is, $H'=\mathrm{d}H/\mathrm{d}N$ and
$H''=\mathrm{d}^2H/\mathrm{d}N^2$. In order to simplify the
mathematical expressions, we introduce the function $G(N)=H^2(N)$,
in terms of which, Eq. (\ref{newfrw1}), can be written as follows,
 \begin{align}\label{newfrw1modfrom}
& -9G(N(R))\left ( 4G'(N(R))+G''(N(R)) \right )F''(R)
\\ \notag & +\left (3G(N)+\frac{3}{2}G'(N(R)) \right
)F'(R)-\frac{F(R)}{2}=0\, ,
\end{align}
where $G'(N)=\mathrm{d}G(N)/\mathrm{d}N$ and also
$G''(N)=\mathrm{d}^2G(N)/\mathrm{d}N^2$. Also, the Ricci scalar can
be written as a function of $G(N)$ as follows,
\begin{equation}\label{riccinrelat}
R=3G'(N)+12G(N)\, .
\end{equation}
This is basically the reconstruction method we shall use, and by
having the scale factor and the Hubble rate at hand, upon combining
Eqs.  (\ref{efoldpoar}) and (\ref{riccinrelat}), we can solve the
second order differential equation (\ref{newfrw1modfrom}), with
respect to the function $F(R)$, thus finding the analytic form (if
possible) of the $F(R)$ gravity which generates the evolution which
is given in terms of $a(t)$ and $H(t)$.

So for the stiff matter era evolution of Eq. (\ref{stiffevo}), let
us express the function $G=H^2$, in terms of the scale factor, which
is,
\begin{equation}\label{hpscf}
G(N)=H^2=\gamma  e^{-6 N}\, .
\end{equation}
Notice that in the Einstein gravity description, $H^2\sim \rho \sim
a^{-6}$ for the stiff matter era. However, in the case at hand,
there is no matter, we
still however, want to stress that Eq. (\ref{hpscf}) implies, $H^2\sim a^{-6}$, since $\ln a=N$,
which is similar with Einstein-Hilbert gravity. Moreover from Eq. (\ref{JGRG11B}), it can be seen that $\rho_\mathrm{eff}\sim H^2$, so in a generalized way, even in the vacuum $F(R)$ gravity case, the total effective energy density satisfies $\rho_\mathrm{eff}\sim a^{-6}$, without of course requiring the presence of stiff matter, and $\rho_{\mathrm{eff}}$ depends solely on the $F(R)$ gravity, see Eq. (\ref{Cr4}).

Also
$\beta$ in Eq. (\ref{hpscf}) is a constant parameter which has a
dependence on $a_0$ and $A$, but we keep a simplified notation for
all these constant, so we use $\gamma$. By combining Eqs.
(\ref{riccinrelat}) and (\ref{hpscf}) we can have the e-foldings
number $N$ as a function of the Ricci scalar $R$, in the following
way,
\begin{equation}\label{efoldr}
N=\frac{1}{6} \ln \left(-\frac{6 \gamma }{R}\right)\, .
\end{equation}
Substituting $G(N)$ from Eq. (\ref{hpscf}) in the differential
equation Eq. (\ref{newfrw1modfrom}) and also by using
(\ref{efoldr}), the final differential equation which can determine
the resulting $F(R)$ gravity is,
\begin{equation}\label{diffeqnforstifffr}
-3 R^2 F''(R)+R F'(R)-\frac{F(R)}{2}=0\, ,
\end{equation}
which can be solved analytically and the solution is,
\begin{equation}\label{frstiff}
F(R)=\beta R^{\mu}+\delta R^{\nu}\, ,
\end{equation}
where $\mu$ and $\nu$ stand for,
\begin{equation}\label{muandnu}
\mu=\frac{1}{6} \left(4+\sqrt{10}\right),\,\,\,\nu=\frac{1}{6}
\left(4-\sqrt{10}\right)\, .
\end{equation}
The parameters $\beta$ and $\delta$ are arbitrary parameters, and in
the following sections we will show that these can be constrained by
the process of gravitational baryogenesis, by the primordial
gravitational waves observational constraints and also by the
inflationary era in the model we describe in the next section.
Clearly, the term $R^{\mu}$ may dominate at some early phase
(depending on the model one uses) of the Universe, when $t\sim 1$sec
or possibly earlier, depending on the free parameters choices, but
we shall discuss this issue further later on.

\section{A Phenomenologically Appealing Combined Patch-$F(R)$ Gravity Model}

In this section we shall introduce a qualitative description of an
$F(R)$ gravity model that may describe the early-time evolution of
the Universe until the Big Bang nucleosynthesis epoch. The new
feature that this model introduces is the existence of a stiff era
generated by an appropriate $F(R)$ gravity term, which we found in
the previous section. In the context of this model, the stiff era
comes after the inflationary era. The functional form of the $F(R)$
gravity is of the following form,
\begin{equation}\label{stifffrearly}
F(R)=R+\frac{R^2}{36H_i}+\beta R^{\mu}+\delta R^{\nu}\, ,
\end{equation}
where the parameters $\mu$ and $\nu$ are given in Eq.
(\ref{muandnu}), and $H_0$ and $H_i$ are parameters which will be
fixed by the Planck constraints on the inflationary era, see the
review \cite{reviews1} for details. The model of Eq.
(\ref{stifffrearly}) may have interesting phenomenology, since the
model (\ref{stifffrearly}) may be appropriately adjusted in terms of
its parameters, so that the $R^2$ term drives the inflationary era,
and it dominates until the term $R^{\mu}$ starts to dominate. This
effect could be appropriately adjusted to occur for cosmic times of
the order $\mathcal{O}(10^{-13})$sec and beyond (which
certain Grand Unified Theories predict that at $t\sim 10^{-13}$sec
the inflationary era ends) as we show shortly. In effect we have a
vacuum $R^2$ gravity dominating and determining the evolution at
early times and for large curvatures, and as the curvature drops,
the evolution starts to be dominated by the stiff matter era
generator $R^{\mu}$. In order to be as concrete as possible, we
shall present all the qualitative details of the model at hand. At
large curvatures, only the first two terms of the model dominate, in
which case the evolution is a quasi-de Sitter one, of the form
\cite{reviews1},
\begin{equation}
\label{quasievolbnexampler2} H(t)\simeq H_0-H_i(t-t_k)\, .
\end{equation}
where $t_k$ is the time instance that the horizon crossing occurs.
It is important to have a concrete idea on when the inflationary era
ends, so this occurs when the first slow-roll parameter becomes of
the order $\epsilon_1\simeq \mathcal{O}(1)$. Suppose that the Hubble
rate at $t=t_f$ is, $H(t_f)=H_f$, where $t_f$ is the time instance
that inflation ends. Then, the condition $\epsilon_1(t_f)\simeq 1$,
yields, $H_f\simeq \sqrt{H_i}$, and in effect we have,
\begin{equation}
\label{timerelation}
t_f-t_k=\frac{H_0}{H_i}-\frac{\sqrt{H_i}}{H_i}\, .
\end{equation}
which at leading order is approximately,
\begin{equation}
\label{timerelation1} t_f-t_k\simeq \frac{H_0}{H_i}\, .
\end{equation}
By using the results above, the spectral index of the primordial
curvature perturbations $n_s$ and the scalar-to-tensor ratio $r$,
can be calculated and these are found to be equal to
\cite{reviews1,Odintsov:2015gba},
\begin{equation}
\label{spectrscatotensor} n_s\simeq 1-\frac{4 H_i}{\left(H_0-\frac{2
H_i N}{H_0}\right)^2}\, , \quad r=\frac{48 H_i^2}{\left(H_0-\frac{2
H_i N}{H_0}\right)^4}\, .
\end{equation}
We can further proceed with the observational indices, by expressing
the $e$-foldings number $N$ in terms of the parameters $H_i$ and
$H_0$, by using, so at leading order we get,
\begin{equation}
\label{nfinal1} N=\frac{H_0^2}{2H_i}\, .
\end{equation}
Now by taking the large-$N$ limit of the indices
(\ref{spectrscatotensor}), we have the standard result,
\begin{equation}
\label{jordanframeattract} n_s\simeq 1-\frac{2}{N}\, ,\quad r\simeq
\frac{12}{N^2}\, .
\end{equation}
From Eq. (\ref{nfinal1}) and for $N=60$, we have that
$\frac{H_0^2}{2H_i}\sim 60$ at leading order. Also, from standard
results \cite{reviews1,Odintsov:2015gba}, it is known that $H_0\sim
10^{13}$sec$^{-1}$, therefore $H_i\sim 8.33333\times
10^{23}$sec$^{-2}$. These values will determine the constraints on
the parameters $\beta$ and $\delta$, in order for the $R^2$ gravity
to dominate during the inflationary era.

Due to relation (\ref{timerelation1}), in this model the
inflationary era ends around $t_f\sim 1.25\times 10^{-11}$sec. After
that, if the third term in Eq. (\ref{stifffrearly}) was absent, the
reheating era would start immediately after the inflationary era.
However, in the case at hand, the third term affects the evolution,
and alters the phenomenology as we demonstrate in later sections.
The exact time instance that the term $R^{\mu}$ will start to
dominate the evolution, cannot be found in an analytic way, since we
would need to solve the cosmological equations analytically, for the
$F(R)$ gravity of Eq. (\ref{stifffrearly}), but we can have a rough
estimate by using the fact that when the term $R^{\mu}$ dominates
the evolution, the scale factor will have approximately the form
$a(t)\sim t^{1/3}$.

By equating the $R^2$ and the $R^{\mu}$ terms by using the
inflationary quasi-de Sitter evolution (\ref{quasievolbnexampler2}),
we can find the constraints on $\beta$ and $\delta$, in order the
$R^2$ dominates until the very last stages of inflation. By doing
so, we obtain at leading order,
\begin{equation}\label{archetyp}
t_s\simeq \frac{4 H_0^4-\beta  H_i 12^{\mu } \left(H_0^2\right)^{\mu
}}{16 H_0^3 H_i}
\end{equation}
where $t_s$ is the time instance that the $R^2$ and the $R^{\mu}$
terms in Eq. (\ref{stifffrearly}) become of the same order. The
relation (\ref{archetyp}) leaves enough space for model building and
for providing interesting phenomenology and also the parameters
$\beta$ and $\delta$ can be constrained.

For example, by appropriately fixing the parameter $\beta$, the time
instance that the stiff era commences may vary. The parameter $H_i$
is constrained by the $R^2$ inflation era, and as was shown in
\cite{Odintsov:2015gba}, it must be approximately $H_i\sim
8.333\times 10^{23}$sec$^{-2}$. Before going into that issue, we
need to stress that the parameter $\beta$ will be further
constrained by the gravitational baryogenesis procedure and also
from the gravitational waves, as we show at a later section.

There are two phenomenologically interesting scenarios which we
shall discuss. In the first scenario, which is more plausible, the
stiff era commences exactly after the inflationary and prior to the
radiation domination era, so for approximately $t_s\simeq
10^{-13}$sec. From Eq. (\ref{archetyp}), this would imply that
$\beta\simeq 0.000219659$sec$^{2\mu-2}$.

Now let us find the constraints on $\beta$, which must be imposed by
the condition that during inflation, the $R^2$ term must dominate
the evolution. By using $R(t)=12H(t)^2+6\dot{H}$, and for the
evolution (\ref{quasievolbnexampler2}) we get at leading order in
the cosmic time that $R\sim \frac{239
H_0^2}{20}-\frac{H_0^3 t}{5}$. This would imply that for
the value of $H_i$ we mentioned earlier, namely  $H_i\sim
8.33333\times 10^{23}$sec$^{-2}$, we would have that the $R^2$ term
is approximately equal to,
\begin{equation}\label{newtermlog1}
\frac{R^2}{36H_i}\simeq \frac{57121 H_0^2}{120}-\frac{239
H_0^3 t}{15}\, ,
\end{equation}
and the $R^{\mu}$ term,
\begin{equation}\label{rmucons}
\beta R^{\mu}\simeq \beta  \left(\frac{239}{20}\right)^{\mu }
H_0^{2 \mu }\, .
\end{equation}
Hence, by requiring $\frac{R^2}{36H_i}>\beta R^{\mu}$ for all cosmic
times in the range $t\sim [10^{-35},10^{-13}]$sec, we have,
\begin{equation}\label{betaconstraint}
\beta<\frac{\frac{57121 H_0^2}{120}-\frac{239 H_0^3
t}{15}}{\left(12 \left(H_0-\frac{H_0^2
t}{120}\right)^2-\frac{H_0^2}{20}\right)^{\mu }}\, .
\end{equation}
By choosing the range $t\sim [10^{-35},10^{-13}]$sec, we obtain that
the constraint of the inflationary era on $\beta$, is,
\begin{equation}\label{constraintbeta}
\beta <0.000223325\mathrm{sec}^{2\mu-2}\, .
\end{equation}
Notice that the value of $\beta$ we found earlier, namely
$\beta\simeq 0.000219659$sec$^{2\mu-2}$ (see above Eq.
(\ref{newtermlog1})), satisfies this constraint.

By adopting the same procedure, the constraint of the inflationary
era on $\delta$ is,
\begin{equation}\label{constraintdelta}
\delta<7.64\times 10^{24}\mathrm{sec}^{2\nu-2}\, .
\end{equation}
Also by requiring $\beta R^{\mu}>\delta R^{\nu}$ up to $t\sim 1$sec,
we further obtain that,
\begin{equation}\label{constraintonbdelta}
\beta>\frac{\delta}{10^{51}}\, ,
\end{equation}
which further constraints $\delta$, and actually it seems that
$\delta$ must take particularly small values.

Hence, according to the scenario at hand, if the constraints
(\ref{constraintbeta}), (\ref{constraintdelta}) and
(\ref{constraintonbdelta}) hold true, the Universe expands after the
inflationary era, in a faster rate, in comparison to the radiation
dominated era, and this is plausible for the simple reason that this
scenario leaves room for enough reheating after the end of the stiff
era. Hence, the stiff $F(R)$ gravity era occurs before the
nucleosynthesis and in the context of our model, there is the
possibility that prior to nucleosynthesis, we may have a stiff
$F(R)$ gravity era, governed by an $R^{\mu}$ term. In effect, this
would possibly have implications on the baryogenesis process, since
it is believed that the later occurs at some time during the
electroweak era. In the literature the presence of a stiff matter
era, indeed may affect the baryogenesis mechanism, but in the
context of sphaleron field configurations \cite{Joyce:1997fc}. In
most cases, the aforementioned problem is addressed in the context
of a canonical scalar field, which after slow-rolling a potential,
it experiences a phase of speeded up expansion, which is called
kination. In the present paper we shall discuss the baryogenesis
issue but in a totally different context, by using solely a vacuum
$F(R)$ gravity. Our main assumption will be that the baryogenesis
actually occurs due to the presence of effective operators, a
mechanism called gravitational baryogenesis
\cite{Davoudiasl:2004gf,Lambiase:2006dq,Odintsov:2016hgc,Oikonomou:2015qfh,
Saaidi:2010ey,Sadjadi:2007dx,Lambiase:2006ft,Bento:2005xk,
Feng:2004mq,Oikonomou:2016jjh,Oikonomou:2016pnq,
Odintsov:2016apy,Arbuzova:2016cem,Lima:2016cbh,Arbuzova:2017vdj}. In
this way, the Universe after the inflationary phase, undergoes an
evolution with scale factor $a(t)\sim t^{1/3}$, for cosmic times
prior to nucleosynthesis and of the order $t\sim
\mathcal{O}(10^{-13})$sec, which is faster than radiation.

According to the other scenario, the stiff $F(R)$ gravity era may
occur exactly before the nucleosynthesis era, so for $t_s\sim
\mathcal{O}(1)$sec. However in this scenario, nucleosynthesis will
start with the Universe being in a non-radiation dominated phase,
which is less appealing in comparison to the former scenario. In the
rest of this paper we focus on the first scenario.

Before we close this section, it is worth mentioning that it is possible to reformulate the $F(R)$ terms in such a way that the parameters $\beta$ and $\delta$ are dimensionless. For instance, the $R^{\mu}$ term can be written,
\[ \beta R^\mu
= \tilde\beta R_a (R/R_a)^\mu, \] 
where $R_a$ is some curvature scale,
so that $\tilde\beta$ is dimensionless. Intuitively, one may choose
the inflationary scale to be $R_a \sim H_{inf}^2$, but this is not the only
choice of course. This parametrization might give us some hint on
the ``naturalness'' of the model considered in this section, for instance if we find
that $\tilde\beta$ must be extremely small. Then the constraints on these
dimensionless parameters would have a clearer interpretation than
constraints on $\beta$ and $\delta$, and the resulting picture can easily be worked out, but we omit this task for brevity.

\subsection{Gravitational $F(R)$ Gravity Baryogenesis During the Stiff Era}

In most contexts, baryogenesis is assumed to occur during the
electroweak phase of the Universe's evolution, but also in some
alternative approaches, baryogenesis may occur at the Grand Unified
Theory (GUT) mass scale. In the context of electroweak baryogenesis,
the baryon asymmetry occurs during the radiation domination era,
which initiates after the end of inflation and during the reheating
process.

A  baryon asymmetry generating mechanism, alternative to usual
electroweak theory approaches, has appeared some time ago, and it is
known as gravitational baryogenesis
\cite{Davoudiasl:2004gf,Lambiase:2006dq,Odintsov:2016hgc,Oikonomou:2015qfh,
Saaidi:2010ey,Sadjadi:2007dx,Lambiase:2006ft,Bento:2005xk,
Feng:2004mq,Oikonomou:2016jjh,Oikonomou:2016pnq,
Odintsov:2016apy,Arbuzova:2016cem,Lima:2016cbh,Arbuzova:2017vdj},
which may provide insights and hints on the excess of matter over
antimatter. The aforementioned excess is supported by  observational
data coming from the Cosmic Microwave Background, so it is
considered as a realistic phenomenological problems of contemporary
physics, and in fact one of the most demanding problems to solve.
Constructed in such a way that one of the Sakharov criteria
\cite{sakharov} is satisfied, gravitational baryogenesis might
explain the baryon-anti-baryon asymmetry, by the presence of an
$\mathcal{C}\mathcal{P}$-violating effective operator, which has the
following form,
\begin{equation}\label{baryonassterm}
\frac{1}{M_*^2}\int \mathrm{d}^4x\sqrt{-g}(\partial_{\mu} R)
J^{\mu}\, .
\end{equation}
The effective operator (\ref{baryonassterm}), may occur from higher
order interactions originating from the higher scale effective
theory which controls the GUT scale physics. The parameter $M_*$ is
characteristic of this underlying effective theory and it stands for
the cutoff scale of this theory. In addition, the current $J^{\mu}$
is the baryonic matter fermion current, while $g$ and $R$ are the
trace of the metric tensor of the geometric background, which will
be assumed to be the FRW one of Eq. (\ref{metricfrw}), and the Ricci
scalar respectively. This means that, for a flat FRW Universe, the
resulting the baryon to entropy ratio, which we denote $\eta_B/s$,
is effectively $\eta_B/s\sim \dot{R}$. In the ordinary approach of
Ref. \cite{Davoudiasl:2004gf}, the baryon to entropy ratio
$\eta_B/s$ for an Einstein-Hilbert Universe is zero, if it is
calculated for a radiation dominated Universe, since the resulting
Einstein equations are,
\begin{equation}\label{ricci}
R=-8\pi G (1-3w)\rho\, .
\end{equation}
where $\rho$ is the energy density of the matter fluid present, with
equation of state $p=w\rho$. In effect, the quantity $\dot{R}$
reads,
\begin{equation}\label{newdotr}
\dot{R}=\sqrt{3}(1-3w)(1+w)\frac{\rho^{3/2}}{M_p^2}\, .
\end{equation}
Due to the fact that $\eta_B/s\sim \dot{R}$, the predicted
baryon-to-entropy ratio is zero for a radiation dominated Universe.

In the present paper we shall make the crucial assumption that the
gravitational baryogenesis takes place during the stiff $F(R)$
gravity era, in which case the scale factor behaves as $a(t)\sim
t^{1/3}$ and the evolution is generated by the $F(R)$ gravity term
of the form $F(R)\sim \beta R^{\mu}$. By taking into account the
results of the previous section, in our description, the
gravitational baryogenesis mechanism takes place after $t\sim
10^{-13}$sec, and for the all time that the stiff era lasts. The
observational constraint on the baryon to entropy ratio is
$\frac{\eta_B}{s}\preceq 9.2\times 10^{-11}$, so we shall calculate
the baryon to entropy ratio for the $F(R)$ gravity at hand, and we
will investigate how this observational constraint may restrict the
free parameters of the theory, namely $\beta$.

Before discussing the $F(R)$ gravity case, let us first study in
brief what would happen in the Einstein-Hilbert case, for a stiff
matter era. In most cases studied in the literature, the stiff
matter era equation of state is $p=\rho$, so $w=1$, and by
substituting in Eq. (\ref{newdotr}), we obtain,
\begin{equation}\label{newdotr111}
\dot{R}=-4\sqrt{3}\frac{\rho^{3/2}}{M_p^2}\, ,
\end{equation}
which is non-zero. Hence, even when the stiff era occurs in the
context of Einstein-Hilbert gravity, the gravitational baryogenesis
mechanism yields a non-zero baryon-to-entropy ratio, in contrast to
a radiation dominated era, which yields $\frac{\eta_B}{s}=0$.

Let us now proceed to the $F(R)$ gravity case, and as the Universe's
expands, the temperature of the Universe as a whole drops.

However, the concept of temperature in the context of vacuum $F(R)$ gravity may be vague, since we assumed that no matter fluids are present. The correct assumption is that matter does not affect the evolution, hence particles generated and contributing to the Grand Unified Theory that governs the particle spectrum, can be present. These however do not drive the evolution, or at least these have subleading effects on the Universe's evolution during the inflation era, and near the end of the inflationary era. Hence, the temperature can be defined due to the fact that particles are present. In the ordinary Starobinsky model, when the term $\ddot{H}$ is not negligible anymore, the slow-roll era ends and the reheating process commences, in which the matter fields that are present will be eventually excited, and the cold and vast Universe will be reheated. In the simplest case, if the matter fields are quantified in terms of a scalar field $\phi$, which satisfies $g^{\mu \nu}\phi_{;\mu \nu}=0$, the curvature contributes to the energy density of this scalar field, in terms of the averaged square of the curvature as follows \cite{reviews1},
\begin{equation}
\label{liatdiff}
\frac{ d \rho}{ d t}=-4\rho H+\frac{\omega
\bar{R}^2}{1152\pi}\, .
\end{equation}
The situation for the stiff era baryogenesis is similar, since the matter fluids do not contribute to the baryogenesis mechanism, or the reheating mechanism, until the effective operator of Eq. (\ref{baryonassterm}) starts to become active and affects the baryogenesis, which occurs at the mass scale $M_*$. At that point, the matter fluids $J^{\mu}$ in Eq. (\ref{baryonassterm}), contribute to the baryogenesis, and the gravitational baryogenesis mechanism takes place. It is conceivable that although before that occurs, the particles did not affect the evolution, the temperature of the Universe could be defined, since the particles were present.

Having clarified this important issue, we proceed and we shall
assume that the gravitational baryogenesis takes place at a critical
temperature $T_D$, which corresponds to the stiff era which occurs
at the time instance $t\sim 10^{-13}$sec. At that point the
Universe's temperature is approximately $T\sim 10^{18}$K, which
corresponds to $T\sim 10^5$GeV. Therefore we shall assume that the
critical temperature is $T_D=10^5$GeV. As the Universe evolves in a
stiff $F(R)$ gravity way, the temperature drops below $T_D$, and a
net baryon asymmetry remains in the Universe, which is equal to,
\begin{equation}\label{baryontoentropy}
\frac{n_B}{s}\simeq -\frac{15g_b}{4\pi^2g_*}\frac{\dot{R}}{M_*^2
T}\Big{|}_{T_D}\, ,
\end{equation}
with $g_b$ being the total number of the baryonic degrees of
freedom, $g_*$ stands for the total number of the degrees of freedom
corresponding to the massless particles. Following the method of
Ref. \cite{Lambiase:2006dq}, by using Eqs. (\ref{stiffevo}),
(\ref{baryontoentropy}) and also that $F(R)\sim \beta R^{\mu}$, the
resulting baryon to entropy ratio is found to be,
\begin{equation}\label{resultbaryontoentropy}
\frac{n_B}{s}\simeq \frac{T_D^{\frac{6}{\mu }-1} \left(g_b
5^{1-\frac{3}{2 \mu }} \pi ^{\frac{3}{\mu }-2} \left(-\frac{g_*
2^{-\mu -1} 3^{\mu -1}}{\beta  (\mu  (3 \mu -4)-1)
M_p^2}\right)^{\frac{3}{2 \mu }}\right)}{M_*^2}\, .
\end{equation}
By using the constraint $\frac{\eta_B}{s}\preceq 9.2\times
10^{-11}$, we can restrict the values of the parameter $\beta$. In
principle, the presence of $M_*$, $g_b$, $g_*$ and $T_D$, offers
room for interesting phenomenological scenarios.
\begin{figure}[h]
\centering
\includegraphics[width=20pc]{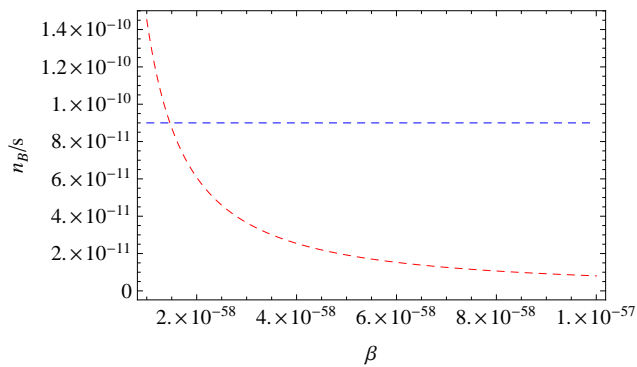}
\caption{{\it{The $\beta$-dependence of the baryon-to-entropy ratio
$n_B/s$ of Eq. (\ref{resultbaryontoentropy}), for $M_*=10^{12}$GeV,
$g_b\simeq \mathcal{O}(1)$, $T_D=10^5$GeV and $g_*\simeq 106$. The
blue line denotes the limiting case $\frac{\eta_B}{s}\simeq
9.2\times 10^{-11}$.}}} \label{plot1}
\end{figure}
Let us investigate how $\beta$ is constrained for a specific choice
of the parameters, so assume that $M_*=10^{12}$GeV, $g_b\simeq
\mathcal{O}(1)$, and $g_*\simeq 106$, and in Fig. \ref{plot1} we
plotted the behavior of the baryon to entropy ratio as a function of
$\beta$, by also taking into account that $T_D=10^5$GeV. For the
choices of the free parameters that we made, the following
constraint on $\beta$ is obtained,
\begin{equation}\label{betagravbaryogenconstr}
\beta > 6.83039 \times 10^{-34}\mathrm{sec}^{2\mu-2}\, ,
\end{equation}
which obvious is satisfied for the choice of $\beta$ we made in the
previous section.

Hence in this section, we investigated the gravitational
baryogenesis scenario in the context of a stiff era generated by an
$F(R)$ gravity. In principle, there are many different scenarios
that can be adopted, but we presented one of these for illustrative
reasons. Along with gravitational baryogenesis, the observational
constraints on primordial gravitational waves can further restrict
the parameter $\beta$. This is the subject of the next section.

\section{Gravitational Waves Constraints on the $F(R)$ Gravity Stiff Era}

Gravitational waves and especially the primordial ones, could be
potentially the direct validation of a theoretical description on
the large scale structure, or at least the gravitational waves can
constrain the existing theories. The primordial gravitational waves
generate a stochastic background in the Universe, the back-reaction
of which can be detected in the near future. However, not only
primordial gravitational waves contribute to the stochastic
background, since there are also astrophysical sources of
gravitational waves, which of course occur in more recent epochs. It
is then possible that astrophysical gravitational waves may dominate
over the primordial ones, however for the particular cases of
modified gravities, there are extra degrees of freedom that may act
as source of specific modes of the primordial gravitational waves.
In this way, the primordial gravitational waves may be
distinguished, and also the modified gravities may be directly
tested. For a recent account on these issues, focusing on the effect
of extended gravities on gravitational waves, see
\cite{Bogdanos:2009tn} and also
\cite{Capozziello:2008fn,Capozziello:2008rq,Ananda:2007xh,Corda:2010zza,Zhang:2014kla,Yang:2011cp,Vainio:2016qas,Bellucci:2008jt}
for $F(R)$ gravity studies on gravitational waves. In the case that
the modified gravity is an $F(R)$ gravity, apart from the tensorial
degrees of freedom of a gravitational wave, there exist an extra
scalar-tensor degree of freedom, the detection of which can validate
that an $F(R)$ gravity is the correct description of the primordial
Universe. The stochastic background of scalar gravitational waves
can be quantified in terms of a scalar field and it is characterized
by a dimensionless spectrum. The full analysis on this topic was
performed in Ref. \cite{Capozziello:2008fn}, so at this point we
shall use the main results in order to constrain the stiff era
$F(R)$ gravity.

Our main assumption is that the primordial gravitational waves are
generated during the stiff era evolution. In the literature, there
exist works that also studied this issue, but in the context of
Einstein-Hilbert \cite{Giovannini:1998bp}. As it was shown in
\cite{Giovannini:1998bp}, if the gravitational waves were generated
during the stiff epoch, this would lead to an overall increase of
the graviton frequency, an effect which would act as a
thermalization mechanism, alternative to reheating. However in this
work we shall be interested on the scalar mode of the gravitational
waves, and the full study of gravitational waves during the stiff
epoch will be presented in a future work.

The motivation to assume that the gravitational waves may be
generated during the stiff era, comes mainly from the fact that
prior to nucleosynthesis, the equation of state of the Universe, and
the general thermodynamical state, is still unknown. Therefore, the
purpose of this section is to study the implications of the
primordial gravitational waves constraints, on the stiff era $F(R)$
gravity, which is $F(R)\sim \beta R^{\mu}$. As we will show in this
section, the constraints coming from the LIGO collaboration,
constrain the allowed values of the parameter $\beta$.

Let us describe in brief the scalar mode of the gravitational waves
in the context of $F(R)$ gravity, following Ref.
\cite{Capozziello:2008fn}. By using the conformal time, the metric
of the geometric background is,
\begin{equation}\label{meetricgw}
\mathrm{d}s^2=a^2(\tau)\left[-\mathrm{d}\tau^2+\mathrm{d}\vec{x}^2+h_{\mu
\nu}(\tau,\vec{x})\mathrm{d}x^{\mu}\mathrm{d}x^{\nu} \right]\, ,
\end{equation}
with the term $h_{\mu \nu}$ describing the gravitational wave. The
tensor $h_{\mu \nu}$ can be decomposed into tensor and scalar modes
in the $z+$ direction as follows,
\begin{equation}\label{ztdependece}
h_{\mu \nu}(t-z)=A^+(t-z)e_{\mu \nu}^++A^x(t-z)e_{\mu \nu}^x+\Phi
(t-z)e_{\mu \nu}^s\, ,
\end{equation}
and the pure scalar mode is the last one, namely,
\begin{equation}\label{cryingintheshadows}
\bar{h}_{\mu\nu}=\Phi e_{\mu \nu}^s\, .
\end{equation}
The LIGO constraint on the scalar mode for a frequency of $100$Hz,
is $\Phi (100Hz)<2 \times 10^{-26}$, so let us now investigate how
this can constrain the stiff $F(R)$ gravity. We can relate the
scalar mode of the gravitational wave to the $F(R)$ gravity in the
Einstein frame as follows,
\begin{equation}\label{einsteinscalarmode}
\Phi\sim  \delta \varphi\, ,
\end{equation}
where $\varphi$ and the corresponding variation are defined as
follows,
\begin{equation}\label{varphi}
\varphi=\ln F'(R),\,\,\, \delta \varphi=\frac{F''(R)}{F'(R)}\delta
R\, ,
\end{equation}
where $R$ is the Ricci scalar. By inserting $F(R)\sim \beta
R^{\mu}$, and by using the LIGO constraint $\Phi <2 \times
10^{-26}$, we obtain,
\begin{equation}\label{ligoconstraintonfr}
-\frac{\beta  (\mu -1) \mu  R_c^{\mu -2}}{\beta  \mu R_c^{\mu
-1}+1}<\frac{2}{10^{26}}\, ,
\end{equation}
where $R_c$ is the curvature during the stiff era. The constraint
(\ref{ligoconstraintonfr}) leads to the following constraint on
$\beta$,
\begin{equation}\label{ligoconstraintfinalonbeta}
\beta >\frac{1}{ \frac{10^{26}}{2}\mu  (\mu -1) R_c^{\mu
-2}+\beta \mu R_c^{\mu-1}}\, ,
\end{equation}
and this the main result of this section. Now, assuming that the stiff era occurs during the time interval $t=[10^{-13},1]$sec, since $a(t)\sim t^{1/3}$ during the stiff era, by using $R(t)=12H^2+6\dot{H}$, we can find approximate expressions for the values of $R_c$, corresponding to $t=1$sec and $t=10^{-13}$sec. For $t=1$sec, $R_c\simeq 0.66$sec$^{-2}$, and for $t=10^{-13}$sec, we obtain $R_c\simeq 6.66\times 10^{25}$sec$^{-2}$. For the value $R_c\simeq 6.66\times 10^{25}$sec$^{-2}$, the constraint (\ref{ligoconstraintfinalonbeta}) results to $\beta>7.27\times 10^{-6}\mathrm{sec}^{2\mu-2}$, while the value $R_c\simeq 0.66$sec$^{-2}$ results to $\beta> 6.18\times 10^{-26}\mathrm{sec}^{2\mu-2}$, so by combining the two constraints, we obtain the final constraint on $\beta$, from the gravitational waves,
\begin{equation}\label{ligofinalcons}
\beta>7.27\times 10^{-6}\mathrm{sec}^{2\mu-2}\, .
\end{equation}
Obviously, the constraint (\ref{ligofinalcons}) is compatible with
the gravitational baryogenesis constraint
(\ref{betagravbaryogenconstr}) of the previous section, and with the
constraint (\ref{constraintbeta}) of section III. Actually, by combining the constraints (\ref{constraintbeta}) and (\ref{ligofinalcons}), we finally find that $\beta$ must be chosen in the following interval,
\begin{equation}\label{constraintbetafinalresultcombined}
7.27\times 10^{-6}\mathrm{sec}^{2\mu-2}<\beta <0.000223325\mathrm{sec}^{2\mu-2}\, .
\end{equation}

\section{Vacuum $F(R)$ Gravity Description of a Stiff and Dust Matter Era}

In this section we shall discuss an alternative topic in comparison
to the previous sections, and specifically we shall investigate
which vacuum $F(R)$ gravity can describe a Universe filled with dust
matter ($w=0$) and stiff matter $w=1$. Thus the content of this
section is different in spirit in comparison to the previous
sections. What we aim to reproduce in the context of vacuum $F(R)$
gravity is the Einstein-Hilbert cosmology determined by the presence
of non-interacting dust matter and also by stiff matter, which may
be some sort of baryonic matter with a stiff equations of state, in
the form of Zel'dovich proposal. This cosmology could describe a
form of matter, which at early times has a stiff equation of state,
and at later times it is described by non-relativistic baryonic
matter. Of course this scenario could be realistic if baryonic cold
non-relativistic matter is taken into account, and in some sense,
this theoretical proposal is a combination of Zel'dovich's early
Universe with the cold dark matter Universe. So the focus in this
section is to realize this by a vacuum $F(R)$ gravity. The
Einstein-Hilbert case was studied in Ref. \cite{Chavanis:2014lra},
and the resulting scale factor of the Universe filled with the
aforementioned perfect fluids, was found to be equal to,
\begin{equation}\label{stiffdustscale}
a(t)=\left(\beta t^2+\gamma t\right)^2\, ,
\end{equation}
where $\beta$ and $\gamma$ are equal to,
\begin{equation}\label{gammaandbeta}
\beta=\frac{9}{4}a_0^3\Omega_{m0}H_0^2,\,\,\,\gamma=3\sqrt{\Omega_{s0}}H_0\,
.
\end{equation}
The corresponding Hubble rate can be written as follows,
\begin{equation}\label{hubbleratedstifdust}
H(t)=\frac{h(t)}{t}\, ,
\end{equation}
where $h(t)$ is equal to,
\begin{equation}\label{htexplicit1}
h(t)=\frac{2 \beta  t^2+\gamma  t}{3 \left(\beta  t^2+\gamma
t\right)}\, .
\end{equation}
As it can be seen by the scale factor (\ref{stiffdustscale}), at
early times it is $a(t)\sim t^{1/3}$ and at later times it scales as
$a(t)\sim t^{2/3}$, so at early times the stiff equation of state
dominates while at later times the dark matter equation of state
dominates. In this section, the focus is to find the vacuum $F(R)$ gravity, which realizes the scale factor (\ref{stiffdustscale}), and not simply the scale factor $a(t)\sim t^{1/3}$, which we studied in the previous section. The difference is that the in the reconstruction of scale factor $a(t)\sim t^{1/3}$ we did in the previous sections, the late-time behavior did not interest us at all. However, the scale factor (\ref{stiffdustscale}) also affects the late-time behavior, for which the scale factor is approximately $a(t)\sim t^{2/3}$. Hence, in the following we shall investigate how the scenario (\ref{stiffdustscale}) may be realized by vacuum $F(R)$ gravity.

To proceed with the calculation, the reason why we wrote the Hubble rate in the form
(\ref{hubbleratedstifdust}), is simply because the function $h(t)$
is a slowly varying function of the cosmic time, and this feature
will simplify the calculations to a great extent, as we will see
later on. It is simple to check that $h(t)$ is a slowly varying
function, since it satisfies,
\begin{equation}\label{r10}
\lim_{t\rightarrow \infty}\frac{h(zt)}{h(t)}=1\, ,
\end{equation}
for all $z$.

In order to find the $F(R)$ gravity description of the stiff and
dust dominated Universe (\ref{hubbleratedstifdust}), we will use a
different reconstruction technique \cite{Nojiri:2009kx}, in
comparison to the one we used in a previous section. The reason is
simply that in this way, we will be able to obtain analytic results.
So we introduce an auxiliary scalar, so that the $F(R)$
gravitational action (\ref{action}) is written as follows,
\begin{equation}\label{neweqn123}
S=\int \mathrm{d}^4x\sqrt{-g}\left ( P(\phi )R+Q(\phi ) \right )
\end{equation}
The two functions $P(\phi )$ and $Q(\phi )$, will essentially
determine the final form of the $F(R)$ gravity, and note that the
auxiliary scalar is identified with the cosmic time, as it was shown
in Ref. \cite{Nojiri:2009kx}. By varying the action
(\ref{neweqn123}) with respect to the auxiliary scalar, we get,
\begin{equation}\label{auxiliaryeqns}
P'(\phi )R+Q'(\phi )=0\, ,
\end{equation}
which when solved with respect to $\phi$, given the functions
$P(\phi)$ and $Q(\phi)$, will yield the function $\phi (R)$. Then
the $F(R)$ gravity can easily be obtained by substituting  $\phi
(R)$ in the $F(R)$ action of Eq. (\ref{neweqn123}), that is,
\begin{equation}\label{r1}
F(\phi( R))= P (\phi (R))R+Q (\phi (R))
\end{equation}
It is then obvious that the main aim of this reconstruction
technique is to determine the functions $P(\phi )$ and $Q(\phi )$.
The differential equation that the aforementioned functions satisfy,
can easily be found by varying Eq. (\ref{neweqn123}) with respect to
the metric, and for a FRW Universe, this becomes,
\begin{align}\label{r2}
& -6H^2P(\phi (t))-Q(\phi (t) )-6H\frac{\mathrm{d}P\left (\phi
(t)\right )}{\mathrm{d}t}=0 \\ \notag & \left ( 4\dot{H}+6H^2 \right
) P(\phi (t))+Q(\phi (t) )+2\frac{\mathrm{d}^2P(\phi (t))}{\mathrm
{d}t^2}+\frac{\mathrm{d}P(\phi (t))}{\mathrm{d}t}=0\, ,
\end{align}
and by eliminating $Q(\phi (t))$ we obtain,
\begin{equation}\label{r3}
2\frac{\mathrm{d}^2P(\phi (t))}{\mathrm {d}t^2}-2H(t) P(\phi
(t))+4\dot{H}\frac{\mathrm{d}P(\phi (t))}{\mathrm{d}t}=0\, .
\end{equation}
The above differential equation can yield the analytic form of the
function $P(\phi)$, given the Hubble rate of the cosmic evolution.
Then, the function $Q(\phi)$ can be found by using the following
formula,
\begin{equation}\label{qfi}
Q(\phi)=-6 H(\phi ) P'(\phi )-6 H(\phi )^2 P(\phi )\, .
\end{equation}
Let us now proceed to the direct calculation of $P(\phi)$ and
$Q(\phi)$ for the case at hand, so by substituting Eq.
(\ref{hubbleratedstifdust}) in Eq. (\ref{r3}), we obtain the
following differential equation,
\begin{align}\label{r12}
& 2\frac{\mathrm{d}^2P(\phi (t))}{\mathrm {d}t^2}-\frac{h(\phi
)}{\phi }\frac{\mathrm{d}P(\phi (t))}{\mathrm{d}t}-\frac{2h(\phi
)}{\phi^2} P(\phi (t)) =0\, ,
\end{align}
where we used the fact that the function $h(t)$ is a slowly varying
function of the cosmic time, so we omitted the higher derivatives
$h'(t),h''(t)$. The differential equation  (\ref{r12}) has the
following general solution,
\begin{align}\label{r16}
& P(\phi )=c_1\phi^{\frac{h(\phi )-1+\sqrt{h(\phi)^2+6h(\phi
)+1}}{2}}+c_2\phi^{\frac{h(\phi )-1-\sqrt{h(\phi)^2+6h(\phi
)+1}}{2}}\, ,
\end{align}
where $c_i$, $i=1,2$ are integration constants. Accordingly, by
substituting the resulting $P(\phi)$ in Eq. (\ref{qfi}), and by also
omitting the higher derivatives of $h(t)$, we obtain the exact form
of $Q(\phi)$,
\begin{align}\label{r18}
& Q(\phi )=-6h(\phi )c_1\left (h(\phi )+\frac{h(\phi
)-1+\sqrt{h(\phi)^2+6h(\phi )+1}}{2}\right )\phi^{\frac{h(\phi
)-1+\sqrt{h(\phi)^2+6h(\phi )+1}}{2}-2}\\ \notag & -6h(\phi
)c_2\left (h(\phi )+\frac{h(\phi )-1-\sqrt{h(\phi)^2+6h(\phi
)+1}}{2}\right )\phi^{\frac{h(\phi )-1+\sqrt{h(\phi)^2+6h(\phi
)+1}}{2}-2}\, .
\end{align}
The fully analytic form of the $F(R)$ gravity can be found in the
two limiting cases, namely, for large $\phi$ (large cosmic times)
and small $\phi$ (small cosmic times). Notice that at large cosmic
times, the scale factor (\ref{stiffdustscale}) behaves as $a(t)\sim
t^{2/3}$, which corresponds to the radiation domination era, and for
small cosmic times, Eq. (\ref{stiffdustscale}) becomes $a(t)\sim
t^{1/3}$, which describes a stiff matter era. We expect that, the
$F(R)$ gravity we found in a previous section, namely $F(R)\sim
R^{\mu}$, will describe the stiff era case. Let us first investigate
the large $\phi$ case, for which $h(\phi)=h_f=\sim \frac{2}{3}$, as
it can be seen from Eq. (\ref{htexplicit1}). In this case, the
function $P(\phi)$ reads,
\begin{equation}\label{pphioneofthefinals}
P(\phi)\sim R^{\frac{h_f-1+\sqrt{h_f^2+6h_f+1}}{4}}\, ,
\end{equation}
so the resulting $F(R)$ gravity reads,
\begin{equation}\label{frgravilast1}
F(R)\sim R^{-\frac{h_f}{2}}\, .
\end{equation}
Accordingly, in the small $\phi$ limit, the function $h(t)$ becomes
$h(t)=h_i=\frac{1}{3}$, hence the function $P(\phi)$ reads,
\begin{equation}\label{pphioneofthefinals1}
P(\phi)\sim \phi ^{\delta }\, ,
\end{equation}
where,
\begin{equation}\label{delta}
\delta=-\sqrt{\frac{2}{3}}
\left(\sqrt{\frac{5}{3}}-\sqrt{\frac{2}{3}}\right)\,.
\end{equation}
Accordingly, the $F(R)$ gravity is,
\begin{equation}\label{frgravityfinal2}
F(R)\sim R^{1-\frac{\delta }{2}}\, .
\end{equation}
The expression in Eq. (\ref{frgravityfinal2}) is identical to
$R^{\mu}$, as it can be checked, with $\mu$ being defined in Eq.
(\ref{muandnu}).

In principle, it is possible to provide a unified description of
inflation, the stiff era (or stiff era with dust) and
radiation-matter domination era, in such a way that there is no
contradiction with the observable predictions of the early Universe,
as we gave hints in the previous sections, and also since the era
after inflation is still mysterious. However, the corresponding
calculation is quite complicated, so we refer from going into
details.

\section{Conclusions}

In this paper we investigated how a stiff matter era can be
generated by a vacuum $F(R)$ gravity, and we discussed several
scenarios which a stiff era can have implications. Firstly, we
demonstrated that the existence of a stiff era following the
inflationary era, may comply with current observational bounds.
Particularly, we considered the phenomenological implications of a
stiff era following an $R^2$ inflationary era. As we showed, by
suitably adjusting the free parameters, it is possible for the stiff
era to occur right after the inflationary era. After discussing the
implications of this requirement on the free parameters of the stiff
$F(R)$ gravity, we investigated how the gravitational baryogenesis
mechanism restricts the stiff $F(R)$ gravity. According to our
approach, a non-zero baryon to entropy ratio can be generated during
the stiff era, both in the $F(R)$ gravity case, but also in the
context of an Einstein-Hilbert gravity. We also discussed how the
scalar mode of the primordial gravitational waves, which is
characteristic to $F(R)$ gravity, may restrict the stiff $F(R)$
gravity, if it is assumed that the gravitational waves are generated
during the stiff era. Finally, we investigated how a combined era
dominated from dust and stiff matter may be realized in the context
of $F(R)$ gravity.

An important issue we did not address,  is the reheating issue.
Specifically, if the stiff era commences after the ending of the
inflationary era, then when and how the reheating era starts? In the
context of $F(R)$ gravity, the reheating may start due to the
existence of intense curvature oscillations, and the $F(R)$ gravity
crucially affects the reheating temperature
\cite{Oikonomou:2017bjx}. Hence, it may be possible that in the
present context, the curvature oscillations during the slow-roll era
may reheat the cold and large Universe, that results after the
extreme accelerating expansion era. Also it is also possible that
the tensorial components of the primordial gravitational waves, if
these are amplified during the stiff era, may reheat the Universe to
a great extent. This scenario actually may occur even in the context
of Einstein-Hilbert gravity, as was shown in Ref.
\cite{Giovannini:1998bp}. It is then possible that this is the case
in the $F(R)$ gravity approach too, that is, high frequency
gravitons may reheat the Universe. However, we need to mention that
the additional reheating that may come from the stiff era will
possibly have a subdominant contribution to the whole reheating
process. We intent to address some of these issues in a future work.

Before closing, let us briefly mention that there exist alternative
scenarios to the one we studied in this paper, in which the stiff
era may occur at a later time and not just after the end of the
inflationary era. It is possible that the stiff era may occur at a
time instance before the nucleosynthesis epoch, so possibly before
$t<1$sec. We did not discuss this scenario, since it is a trivial
extension of the approach we adopted and the resulting phenomenology
would be qualitatively the same, with the scales of the parameters
and temperature being different though.

Finally, it is noteworthy that in principle, our proposal may be
easily extended to other modified gravities, like string-inspired
theories, $F(G)$ gravity, non-local gravity and so on
\cite{reviews1,reviews2,reviews3}.

\section*{Acknowledgments}

This work is supported by MINECO (Spain), project FIS2013-44881,
FIS2016-76363-P and by CSIC I-LINK1019 Project (S.D.O) and by
Ministry of Education and Science of Russia Project No. 3.1386.2017
(S.D.O and V.K.O).

\end{document}